\documentclass[aps,epsf,preprint,nofootinbib]{revtex4}
\usepackage{amsmath}
\usepackage{epsfig}

\newcommand{\nslash}{\kern 0.2 em n\kern -0.50em /}

\newcommand{\beq}{\begin{eqnarray}}
\newcommand{\eeq}{\end{eqnarray}}

\def\bq{\begin{eqnarray}}
\def\eq{\end{eqnarray}}
\def\bq{\begin{equation}}

\def\roughly#1{\mathrel{\raise.3ex\hbox{$#1$\kern-.75em
\lower1ex\hbox{$\sim$}}}}

\begin{document}

\preprint{\hfill\parbox[b]{0.3\hsize}
{ }}

\def\bra{\langle }
\def\ket{\rangle }

\title{
Model calculations of the Sivers function satisfying the Burkardt Sum Rule
}
\author{A. Courtoy$^1$,
S. Scopetta$^{2}$\footnote{corresponding author:
E-mail address: sergio.scopetta@pg.infn.it}, V. Vento$^{1}$}
\affiliation
{\it
$^1$
Departament de Fisica Te\`orica, Universitat de Val\`encia,~and Institut de Fisica Corpuscular, Consejo Superior de Investigaciones Cient\'{\i}ficas, 46100 Burjassot (Val\`encia), Spain
\\
$^2$
Dipartimento di Fisica, Universit\`a degli Studi di Perugia, and INFN sezione di Perugia, via A. Pascoli 06100 Perugia, Italy
}

\begin{abstract}

It is shown that, at variance with previous analyses, the MIT bag model
can explain the available data of
the Sivers function and satisfies the Burkardt Sum Rule
to a few percent accuracy. The agreement
is similar to the one recently found in the constituent
quark model. Therefore, these two model calculations of the Sivers function are in agreement with the
present experimental and theoretical wisdom.

\end{abstract}

\maketitle

The study of 
the partonic properties of transversely polarized hadrons
will answer crucial questions on their structure,
such as their relativistic nature and their angular momentum content. 
Experiments are progressing fast,
motivating a strong theoretical
activity \cite{bdr}.
One of the quantities under scrutiny is the Sivers function,
the object of this study.
Semi-inclusive deep inelastic scattering (SIDIS), i.e.
the process $A(e,e'h)X$, with the
detection in the final state
of a hadron $h$
with the scattered electron $e'$, is one of the
processes to access the transversity parton distributions (PDs).
For several years it has been known that
SIDIS off a transversely polarized target
shows
{the so called ``single spin asymmetries'' (SSAs). 
It can be shown 
(see, i.e., Ref.~\cite{bac1} and references therein),
that one of the mechanisms generating
the SSAs is governed by
the Sivers function \cite{sivers}. The latter
describes the amplitude of modulation of the number density 
of unpolarized quarks
in a transversely polarized target due to the correlation between 
the transverse spin of the target and the intrinsic 
transverse parton momentum.}
The Sivers function is
a Transverse Momentum Dependent (TMD) PD, denoted
$f_{1T}^{\perp \cal{Q}} (x, {k_T} )$, where $x$ is the Bjorken variable
and $k_T$ is the transverse momentum of the parton $\cal Q$.
It is a time reversal odd object \cite{bdr} and
for this reason,
for several years, it was believed to vanish.
However, this argument was invalidated by a calculation
in a spectator model, following the discovery
of Final State Interactions (FSI) at leading-twist, i.e.,
not kinematically suppressed in DIS \cite{brohs}.
The current wisdom is that a non-vanishing Sivers function
is generated by 
FSI, technically represented by the gauge link in the definition of TMD
parton distributions \cite{coll2}.
Recently, the first data
of SIDIS off transversely polarized targets
have shown
a strong flavor dependence of the Sivers mechanism
\cite{hermes}.
Complementary experiments on transversely polarized
$^3$He target,
addressed
in \cite{brod}, are being performed at JLab \cite{06010}.
Parameterizations of $f_{1T}^{\perp \cal{Q}} (x, {k_T} )$
are available \cite{ans,vj,coll3},
and new data are expected soon.
From the theoretical point of view, a model independent constraint
on calculations
of $f_{1T}^{\perp \cal{Q}} (x, {k_T} )$
is the Burkardt Sum Rule (SR)
\cite{Burkardt:2004ur}.
It states that
the average transverse momentum of all the partons in a hadron,
$\langle \vec k_T \rangle$,
which
can be defined through 
$f_{1T}^{\perp \cal{Q}} (x, {k_T} )$,
has to vanish.
If the proton is polarized in the positive $x$ direction,
the Burkardt SR reads:
\begin{equation}
\sum_{{\cal Q}=u,d,s,g..} \langle k_y^{\cal{Q}} \rangle =
- \int_0^1 d x \int d \vec k_T
{k_y^2 \over M}  f_{1T}^{\perp \cal{Q}} (x, {k_T} )=0~.
\label{burs}
\end{equation}
{Given the present situation of increasing experimental activity,  
estimates of $f_{1T}^{\perp \cal{Q}} (x, {k_T} )$,
subject to solid theoretical constraints, can be very useful.}
{Since
a direct calculation in QCD is not yet feasable,
this quantity has been
calculated in several models:}
a quark-diquark model
\cite{brohs,bacch}; the MIT bag model, in its simplest
version \cite{yuan} and introducing an instanton contribution
\cite{d'a};
the Constituent
Quark Model (CQM) \cite{nuestro}.
To distinguish between the model estimates, data
and model independent relations, such as the Burkardt SR, can be used.
In all the models used so far
the total momentum of the proton is carried by the
quarks of flavor $u$ and $d$.  According to Eq. (\ref{burs}),
this implies that the magnitude of $f_{1T}^{\perp \cal{Q}}$ for
${\cal Q}=u$ and $d$ has to be similar and the sign has to be opposite.
This is also the trend of the parameterizations of the data \cite{ans,coll3}.
In different versions
of the diquark model
the magnitude of the $d$ contribution is much
smaller than that of the $u$. In
the MIT bag model \cite{yuan} 
$u$ and $d$-quark contributions of opposite sign are found to be
proportional by a factor of -4.
{Even in the modified version of
the MIT bag model of Ref. \cite{d'a}, the Burkardt SR is not fulfilled. On the
contrary, in the CQM,
we found} 
a satisfactory
description of the data and therefore the calculation fulfills the
Burkardt SR at the 2$\%$ level \cite{nuestro}.
This puzzling situation deserves to be investigated.
To this end,
we next analyze the MIT bag model calculation to understand the origin of the discrepancy with the CQM calculation.
One should realize that, in the CQM, even if a pure S-wave description of the proton is used, i.e., a pure
$SU(6)$ wave function, we are able to reproduce the gross features of the data. The same $SU(6)$ spin-flavor structure is used in the
bag calculation of Refs.~\cite{yuan,d'a} and no agreement with the data is found.
This situation is in contradiction with
previous calculations
of other PDs in the bag model \cite{jaffe} and
in the CQM \cite{trvpg} which both have been able to reproduce  the 
gross features of the data. 

If the proton is polarized in the positive $x$ direction,
the Sivers function can be written, in a helicity basis
for the proton, as
\cite{nuestro,tn}
\begin{eqnarray}
&&f_{1T}^{\perp {\cal Q}} (x, {k_T} ) =
2\,\Re \Big \{
{
M \over 4 k_y
}
\int
{ d \xi^- d^2 \vec{\xi}_T \over (2 \pi)^3 }
e^{-i ( x \xi^- P^+  - \vec{\xi}_T \cdot {{\vec k_T}} )}
\langle P S_z = 1 |
\hat O_{\cal Q}
| P S_z = - 1 \rangle \Big \}~,
\label{work}
\end{eqnarray}
where
$\hat O_{\cal Q} =
\bar \psi_{\cal Q}(0,\xi^-,\vec{\xi}_T)
{\cal L}^\dagger_{\vec{\xi}_T}(\infty,\xi^-)
\gamma^+
{\cal L}_0(\infty,0)
\psi_{\cal Q}(0)~,
$
$\psi_{\cal Q}(\xi)$ is the quark field,
$
{\cal L}_{\vec{\xi}_T}(\infty,\xi^-)
$
is the gauge link \cite{coll3}.
In the following, the framework and the notation of Ref.~\cite{yuan} are used
to calculate Eq. (\ref{work}) in the MIT bag model \cite{bag,jaffe}.
\begin{figure}
\includegraphics[height=4cm]{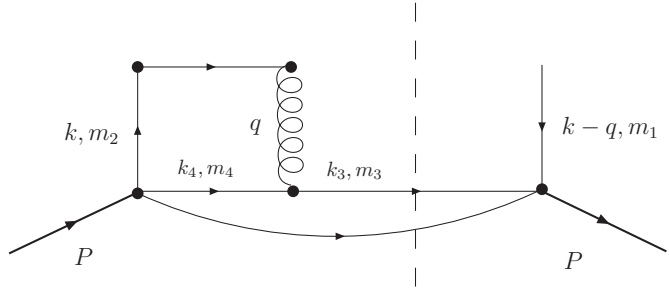}
\caption{The contributions to the Sivers
function in the present approach. The graph has been
drawn using JaxoDraw \cite{Binosi:2003yf}.}
\label{graph}
\end{figure}
By expanding the gauge link to next to leading order, 
inserting in Eq. (\ref{work}) the bag model wave function
in momentum space, $\varphi(k)$ \cite{bag}  , and
using for the definition of the quark 
helicity and momentum labels the ones in Fig.~\ref{graph},
$f_{1T}^{\perp {\cal Q}} (x, {k_T} )$
can be written as
\begin{widetext}
\begin{eqnarray}
f_{1T}^{\perp {\cal Q}}(x,k_\perp)& = &-g^2\frac{ME_P}{k^y}
 \,2\Re \Big\{
\int \frac{d^2q_\perp}{(2\pi)^5}\frac{i}{q^2}
\sum\limits_{\{m\}, \beta}
C_{\{m\}}^{\cal Q,\beta}
\varphi_{m_1}^\dagger(\vec{k}-\vec{q}_\perp)\gamma^0\gamma^+
\varphi_{m_2}(\vec{k}) \Big.
\nonumber
\\
&\times& 
\int\frac{d^3k_3}{(2\pi)^3}\varphi_{m_3}^\dagger(\vec{k_3})\gamma^0\gamma^+
\varphi_{m_4}(\vec{k_3}-\vec{q}_{\perp}) \Big\}~.
\label{siversaur}
\end{eqnarray}
This equation corresponds to Eq. (17) in Ref.~\cite{yuan} modified to follow the Trento convention~\cite{tn}
which implies an additional factor of $1/2$. Here $g$ is the strong coupling constant,
$
C_{\{m\}}^{\cal Q,\beta}
=
T^a_{ij}T^a_{kl}
    \langle PS_z=1 |b_{{\cal Q} m_1}^{i\dagger}b_{{\cal Q} m_2}^{j}
    b_{\beta m_3}^{k\dagger}b_{\beta m_4}^{l} |PS_z=-1\rangle
$ with
$\{m\}=m_1,m_2,m_3,m_4$;
$M$ is the proton mass, $E_p$ its energy, $b_{{\cal Q},m}^i$
is the annihilation operator for a quark with flavor ${\cal Q}$,
helicity $m$, and color index $i$, and $T_{ij}^a$ is a Gell-Mann
matrix. In turn, the $k_3$ integral 
can be written as
\begin{eqnarray}
\int\frac{d^3k_3}{(2\pi)^3}\varphi_{m_3}^\dagger(\vec{k_3})\gamma^0\gamma^+
\varphi_{m_4}(\vec{k_3}-\vec{q}_{\perp})
&\equiv &
 F_{m_3}(\vec{q}_{\perp})\,\delta_{m_3m_4}\,+H_{m_3}
(\vec{q}_{\perp})\,\delta_{m_3,-m_4}~,
\quad  \label{k3}
\end{eqnarray}
with
\begin{eqnarray}
F_{m_3}(\vec{q}_{\perp})
&=&\frac{C}{\sqrt{2}}\,\int d^3k_3
\left[
t_0^3t_0^{'3}+k^z_3 t_1^3t_0^{'3}/k_3+
k_3^{'z}
t_1^{'3}t_0^3/k'^3 \right.
\nonumber
\\
&+&
\left. \big( \vec{k}_3\cdot\vec{k}_3' + i\,v^z d_{m_3}\big)\,
{t_1^3} t_1^{'3} /{ ({k}'_3 k_3)}
\right]~,
\\
H_{m_3}(\vec{q}_{\perp})&=&\frac{C}{\sqrt{2}}\,\int d^3k_3
\left[
\left( i k_3^y - k_3^x d_{m_3}
\right)\,t_0^{'3} {t_1^3}/{k_3}
-\Big( ik_3'^y 
-k_3'^x d_{m_3} 
\Big)\,t_0^3 t_1^{' 3}/{k}'_3 
\right.
\nonumber\\
& + & \left.  \left(v^y d_{m_3}+i\,v^x\right)\,
{t_1^3}
{t_1^{'3}}/
({k}_3
{k}'_3)
\right]~,
\label{us}
\end{eqnarray}
\end{widetext}
where $k_3 = |\vec k_3|$, $k_3' = |\vec k_3'|$,
$d_{m_3}\equiv(\delta_{m_3\frac{1}{2}}-\delta_{m_3,-\frac{1}{2}})$,
$\vec v = \vec{q}_{\perp}\times\vec{k}_3$,
$\vec{k}_3'=\vec{k}_3-\vec{q}$, $C= {16\omega^4}/({\pi^2
j_0^2(\omega)(\omega -1)}{M_P^3})$, with $\omega$ being the
bag model mode \cite{bag} and the function $t_i^3=t_i(
k_3)$, $t_i'^3=t_i(k_3')$, $i=0,1$ are defined in
\cite{yuan}.

In Ref.~\cite{yuan},
 only the first term in the r.h.s. of  Eq.~(\ref{k3})
is calculated.
This term corresponds to the helicity conserving contribution ($m_3=m_4$) associated with the quark 
in the lower part of Fig.~\ref{graph}.
This result only applies if the integral is
performed taking $\vec q_\perp$ along the $z$ direction. However, this
is incorrect in the present case. 
As in any DIS process, the direction of the virtual photon determines the operator structure, i.e. $\gamma^+$ in here.
The fixing of the photon direction leads to a $\gamma_3$ matrix in this operator. 
Therefore we do not have anymore the freedom to choose $z$ 
as the direction of the exchanged gluon,
which must lie in the $(x,y)$ plane.
Besides, one can check that the integral
Eq.~(\ref{k3}) does depend on the direction of $\vec q_\perp$.
Moreover, 
 if the findings of Ref.~\cite{yuan} were correct, it
would mean that a helicity flip could occur only for the quark which interacts with the photon and not for the quark
 in the lower part of Fig.~\ref{graph}, a restriction which does
not  have any physical motivation. 
Thus the present calculation differs from the previous one in that 
we take into consideration both terms of Eq.~(\ref{k3}).
By the same argument, the expression
$\varphi_{m_1}^\dagger(\vec{k}-\vec{q}_\perp)\gamma^0\gamma^+
\varphi_{m_2}(\vec{k})$ 
in Eq. (\ref{siversaur}) 
also contains both helicity-flip and non-flip terms.
{One should notice that, in Ref. \cite{d'a},
a helicity flip term (with $m_3 = -m_4$) has been found to
contribute to the Sivers function. In that paper, instanton
effects have been added to the pure MIT bag model calculation
of Ref. \cite{yuan} and the presence of this helicity flip term
is due solely to the
instanton contribution.
However, in the calculation of Ref. \cite{d'a},
the term depending on $\delta_{m_3-m_4}$ in Eq. (4),
associated to the perturbative
one-gluon exchange, should appear and has not
been considered \cite{pc}.
It is interesting to realize that, in a completely different scenario, 
the CQM calculation satisfying the Burkardt SR of Ref. \cite{nuestro}, 
a contribution is
found for helicity conserving and helicity flip of the quark in the
lower part of Fig 1, as it happens in our MIT bag model calculation.}
\begin{widetext}
Evaluating 
 the matrix
elements for the valence quarks and assuming an $SU(6)$ proton state in Eq. (\ref{siversaur}),
one gets
\begin{figure}
\includegraphics[height=12cm]{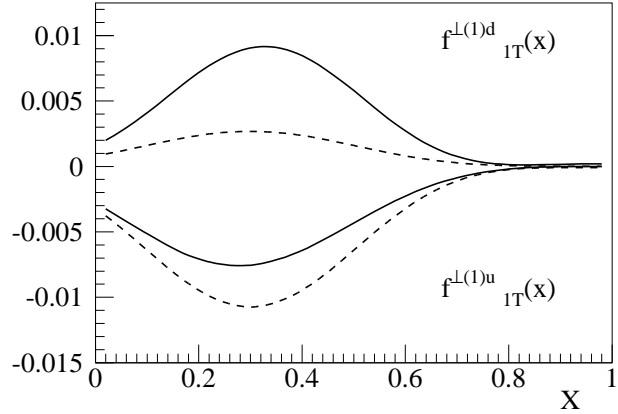}
\vskip -6cm\caption{ The quantity $f_{1T}^{\perp (1)q }(x) $, Eq. (\ref{momf}),
for the $u$ and $d$ flavour.
The dashed curves are the results of the approach
of Ref. \cite{yuan}, the full ones those obtained here.}
\label{filippo}
\end{figure}
\begin{eqnarray}
f_{1T}^{\perp{\cal Q}}(x,k_\perp)&=&-\frac{g^2}{2}\frac{ME_P}{k^y} \,C^2
\int
\frac{d^2q_\perp}{(2\pi)^2}\frac{1}{q^2} [ C_{\cal Q}^{-+} Y(\vec
q_\perp,k_T) + C_{\cal Q}^{+-} U(\vec q_\perp,k_T)]~, \label{sivf}
\end{eqnarray}
with
\begin{eqnarray}
Y(\vec q_\perp,k_T)&=&\left[ k'^y t_1't_0/k'-k_yt_1t_0'/k -
v^y t_1t_1'/(kk')\right]\nonumber\\
&&\times \int d^3k_3 \left[
t_0^3t_0^{'3}+k^z_3t_1^3t_0^{'3}/k_3+k_3^{'z}t_1^{'3}t_0^3/k_3'+\,\left(k_3^2-
\vec{k}_3\cdot\vec{q}_{\perp}\right)\,t_1^3 t_1^{'3}/(k_3
k_3')\right] ~, \label{y}\\
U(\vec q_\perp,k_T)&=&\left[t_0't_0+k^z(t_1't_0/k'+t_0't_1/k) +
(k^2
- \vec q_\perp \cdot \vec k) t_1t_1'/(kk')\right]\nonumber\\
 &&\times 
\int d^3k_3 \left[
 k_3^y\,t_0^{'3}t_1^3/k_3 - k'^y\,t_0^3 t_1^{' 3}/k'_3
+v^x\,t_1^3t_1^{'3}/(k'_3k_3)
 \right]~, \label{u}
\end{eqnarray}
\end{widetext}
where $k = |\vec k| $, $k' = |\vec k'|$,
$t_i=t_i(k)$, $t_i'=t_i(k')$, $i=0,1$
and
$C_u^{-+}=-16/9$ ($C_d^{-+}=4/9$),
$C_u^{+-}=-4/9$ ($C_d^{-+}=-8/9$) for ${\cal Q}=u(d)$.
Let's recall that in Ref. \cite{yuan} only the first term
of Eq. (\ref{sivf}), proportional to $C_{\cal Q}^{-+}$,
contributes to $f_{1T}^{\perp{\cal Q}}$. It is therefore
found that $f_{1T}^{\perp{u}}= -4 f_{1T}^{\perp{d}}$.
Notice that, in order to calculate $f_{1T}^{\perp{\cal Q}}$,
one is using a two-body operator
associated with FSI and therefore one should not expect a proportionality between the $u$ and $d$ results.
On the contrary, in the calculation of conventional PDs, in any SU(6)  model calculation, the used operators
are of one-body type and therefore the results turn out to be proportional \cite{jaffe}.
\begin{figure}[h]
\includegraphics[height=12cm]{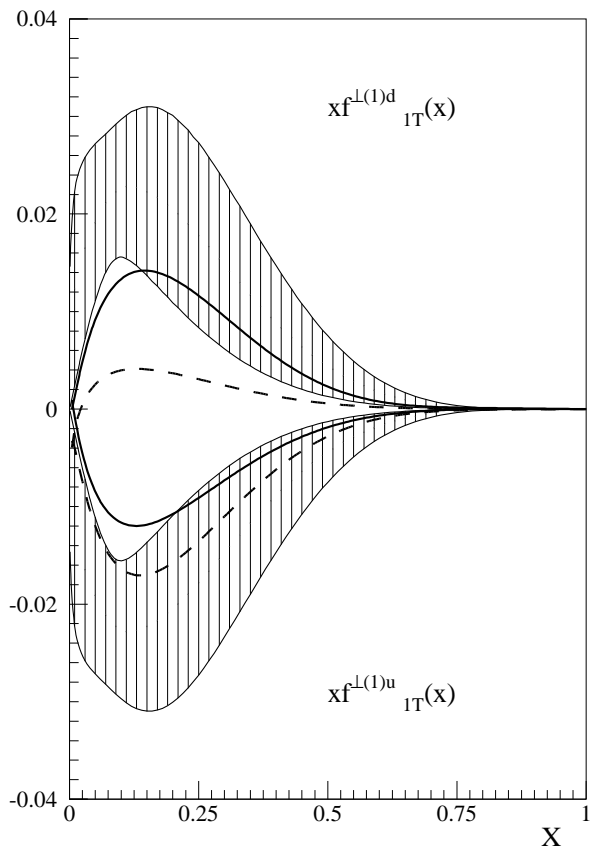}
\caption {The same as in Fig. 2, after
NLO evolution (see text).
The patterned area represents the $1 - \sigma$ range of the best fit
of the HERMES data
proposed in Ref.~\cite{coll3}.}
\label{evo}
\end{figure}
Numerical results are shown in Figs.~\ref{filippo} and~\ref{evo} for the first moment
of $f_{1T}^{\perp{\cal Q}}$, i.e.
\begin{equation}
f_{1T}^{\perp (1) {\cal Q} } (x)
= \int {d^2 \vec k_T}  { k_T^2 \over 2 M^2}
f_{1T}^{\perp {\cal Q}} (x, {k_T} )~.
\label{momf}
\end{equation}
In Fig.~\ref{filippo} the dashed curves are the ones obtained 
in Ref.~\cite{yuan}
(cf. Fig. 4 in Ref \cite{yuan} 
adapted to the Trento convention \cite{tn}, i.e.,
reduced by a factor of two).
The obtained value for the  Burkardt SR Eq.~(\ref{burs}) turns out to be 8.73 MeV.
To have an estimate
of the quality of the agreement of this result with
the SR, we consider the ratio
$r= (
\langle k_x^{d} \rangle+
\langle k_x^{u}\rangle ) /
( \langle k_x^{d} \rangle-
\langle k_x^{u} \rangle )~,$
obtaining $r \simeq 0.60$, i.e., the Burkardt SR seems to be violated by
60 \%.
The full curve in Fig.~\ref{filippo} is the  result
of the present calculation. Clearly, the $d$ contribution becomes comparable
in magnitude to the $u$ one. 
The obtained value for the Burkardt SR is -0.78 MeV 
and $r \simeq 0.05$, i.e., it is only violated by 
5 percent. 
These results are 
comparable in quality to those obtained for the CQM~\cite{nuestro}, restoring
the approximate agreement between the two schemes.
\begin{figure}[h]
\includegraphics[height=12cm]{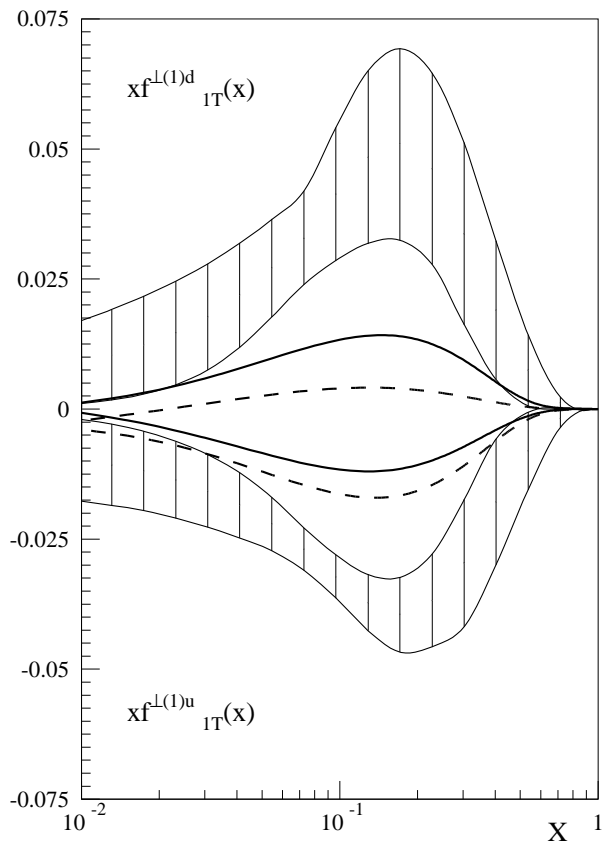}
\caption {The same as in Fig. 3, but comparing with the parameterization
of the data proposed in Ref.~\cite{ans} (patterned area).}
\label{evo1}
\end{figure}
In order to compare the results with the data,
one should perform a QCD evolution from the experimental scale, which
is, for example for the HERMES
data, $Q^2 =2.5$ GeV$^2$~\cite{jaro}. 
Unfortunately, the evolution
of TMDs is still
to be understood, although
recent developements can
be found in Ref.~\cite{cecco}.
In order to have an indication of the effect of the
evolution, we evolve at NLO the model
results assuming, for the moments of the Sivers function,
Eq.~(\ref{momf}),
the same anomalous dimensions of the unpolarized PDFs,
as we did in Ref.~\cite{nuestro} for the CQM calculation.
The parameters of the evolution have been fixed
in order to have a fraction $\simeq 0.55$ of the momentum
carried by the valence quarks at 0.34 GeV$^2$, as in
typical parameterizations of PDs , starting from
a scale of $\mu_0^2 \simeq 0.1$ GeV$^2$ with only valence quarks.
The results in Fig.~\ref{evo} and ~\ref{evo1} show an impressive improvement
of the agreement with data once the full contribution of Eq.~(\ref{sivf})
is taken into account. The
data are described rather well for both flavors.
Comparing this encouraging outcome with that of Ref.~\cite{nuestro},
one can notice that the Burkardt SR is better fulfilled in the CQM.
Most probably this has to do with the fact 
that the Burkardt SR is associated with transverse momentum conservation
and, in the MIT bag
model, the proton wave function is not an exact momentum eigenstate.
In closing, we can say that, for the first time,
it has been established that
correct model calculations provide phenomenological
successful interpretations of the Sivers function,
which are consistent with
each other.

We thank I.O. Cherednikov, A. Drago and
S. Noguera for fruitful discussions,
P. Schweitzer for sending us the parameterization of the data
of Ref.~\cite{coll3},
and A. Prokudin for sending us that of Ref.~\cite{ans}. 
A.C. thanks the Department of Physics of the Perugia
University for warm hospitality.
This work is supported in part by the INFN-CICYT,
by the Generalitat Valenciana, contract
AINV06/118; by the $6^{th}$ FP of the
EC, Contract No. 506078 (I3 Hadron Physics);
by the MEC (Spain), Contract FPA 2007-65748-C02-01 and
grants AP2005-5331 and PR2007-0048.



\end{document}